\documentstyle[preprint,tighten,version2,aps]{revtex}
{\makeatletter%
\gdef\labeleqs#1{{%
\edef\@currentlabel{%
\ifappendixon\appletter\fi
\ifsecnumbers\ifnum\c@secnum>0 
\arabic{secnum}.\fi\fi\arabic{equation}}%
\label{#1}%
}}%
} 
\begin{document}
\input{epsf.tex}
\preprint{IFUP-TH 22/95, SWAT/94-95/76}
\begin{title}
\hskip-10pt Colour confinement as dual Meissner effect: $SU(2)$ gauge theory.
\end{title}
\author{L. Del Debbio\thanks{Address after Jan.~1995: Department of 
Physics, University of Wales, Swansea, Singleton Park, Swansea SA2 8PP, 
U.K.}, A. Di Giacomo,
G. Paffuti, P. Pieri}
\begin{instit}
Dipartimento di Fisica dell'Universit\`a and 
I.N.F.N., I-56126 Pisa, Italy 
\end{instit}
\begin{abstract}
We demonstrate that confinement in $SU(2)$ gauge theory is produced by
dual superconductivity of the vacuum. We show that for $T$ $<$ $T_c$
(temperature of deconfining phase transition) the $U(1)$ symmetry related to
monopole charge conservation is spontaneously broken; for  $T$ $>$ $T_c$
the symmetry is restored.
\end{abstract}
\section{Introduction}
Dual superconductivity of the vacuum has been advocated as the mechanism for
confinement of colour\cite{1,2,3}: the chromoelectric field is channeled into
Abrikosov\cite{4} flux tubes, producing a static potential
proportional to the distance between $q\bar q$ pairs.

Magnetic charges, defined as Dirac monopoles of a residual $U(1)$ symmetry
selected by a suitable gauge fixing (abelian projection), should accordingly
condense in the vacuum, in the same way as Cooper pairs do in an ordinary
superconductor. 

Evidence has been collected by numerical simulations of the theory on the
lattice, that such monopoles do exist, and that their number density is
correlated with the deconfining phase transition: we refer to\cite{5} for a
review of these results. 
For a recent updating we refer to \cite{5a}
However a direct demonstration that confinement is
produced by monopole condensation is still lacking.

In fact monopole condensation means that the ground state of the system is a
superposition of states with different magnetic charges, or that the dual
(magnetic) $U(1)$ symmetry is spontaneously broken, in the same way as the
electric $U(1)$ symmetry is  spontaneously broken in a ordinary
superconductor\cite{6}. Such a breaking is monitored by the non vanishing of
the vacuum expectation value ({\it vev}) of any operator $\mu$ with nontrivial
magnetic charge. $\langle\mu\rangle$ is called a disorder parameter: it is non
zero in the broken phase, and vanishes in the ordered phase, at least in the
thermodynamical limit $V\to \infty$.

The construction of a disorder parameter for dual superconductivity has been
presented in \cite{7} where it has been successfully tested on compact $U(1)$
gauge theory.

In this paper we use the same construction to probe the vacuum condensation of
the monopoles defined by abelian projection in $SU(2)$ gauge theory. We find
that the abelian projection which diagonalizes the Polyakov line\cite{8},
identifies monopoles which condense in the confined phase and do not in the
deconfined one.

Monopoles defined by the abelian projection which diagonalizes a component of
the field strength\cite{8} do not show any signal of condensation correlated
with confinement.

We conclude that:
\begin{itemize}
\item[1)] Confinement of colour is related to dual superconductivity of gauge
theory vacuum.
\item[2)] Not all the abelian projections are equivalent, or define monopoles
wich condense in the vacuum to confine colour.
\end{itemize}

In sect.2 we recall the basic ideas of the abelian projection, the definition
of the corresponding monopoles, and their role in confinement. In sect.3 we
present our results, in sect.4 the conclusions.

\section{Monopoles in Q.C.D.}
Stable monopoles configurations in gauge theories are related to the first
homotopy group $\Pi_1$ of the gauge group\cite{9}. Since $\Pi_1[SU(N)]$ is
trivial, the symmetry has to break down to some non simply connected subgroup,
in order to define magnetic charges.

In the Georgi-Glashow\cite{10} model with gauge group $SU(2)$ coupled to a
scalar field $\Phi^a$ in the adjoint representation a spontaneous breaking
$SU(2)\to U(1)$ allows to define an abelian field strength\cite{11}
\begin{equation}
f_{\mu\nu} = \hat\Phi^a G^a_{\mu\nu} - \frac{1}{g}\varepsilon_{abc}
\hat\Phi^a(D_\mu\hat\Phi)^b(D_\nu\hat\Phi)^c
\label{eq2.1}\end{equation}
which admits stable monopole configuration\cite{11,12}, behaving as Dirac
monopoles at large distances. In Eq.(\ref{eq2.1}) $\hat\Phi^a = \Phi^a/|\Phi|$.
Putting
\begin{equation}
a_\mu = \hat\Phi^a A^a_\mu\label{eq2.2}\end{equation}
one has\cite{13}
\begin{equation}
f_{\mu\nu} = \partial_\mu a_\nu - \partial_\nu a_\mu -\frac{1}{g}
\varepsilon_{abc}
\hat\Phi^a(\partial_\mu\hat\Phi)^b(\partial_\nu\hat\Phi)^c
\label{eq2.3}\end{equation}
$f_{\mu\nu}$ in Eq.(\ref{eq2.1}) is gauge invariant, since both $\hat \Phi$ and
$G_{\mu\nu}$ are gauge covariant. $a_\mu$ in Eq.(\ref{eq2.2}) is not. In the
gauge in which
\begin{equation}
\hat\Phi^a = \delta_3^a \label{eq2.4}\end{equation}
Eq.(\ref{eq2.3}) becomes
\begin{equation}
f_{\mu\nu} = \partial_\mu a_\nu - \partial_\nu a_\mu
\label{eq2.5}\end{equation}
Eq.(\ref{eq2.5}) is the usual expression of the electromagnetic field strength
in terms of the (abelian) potential $a_\mu$. The choice of the gauge 
Eq.(\ref{eq2.4}), which is defined up to an arbitrary gauge rotation around the
third axis, is called an abelian projection.

The strategy for relating confinement of colour to dual superconductivity in
Q.C.D. is to make a guess of a possible effective Higgs field $\hat\Phi$,
belonging to the adjoint representation, and then perform the abelian
projection Eq.(\ref{eq2.4}) and look for condensation of the corresponding
Dirac monopoles.

In most of the lattice investigations on the problem, the density of monopoles,
or quantities related to it, have been studied: of course the density of
magnetic charges is not a disorder parameter for dual superconductivity, in the
same way as the number density of electrons is not for ordinary
superconductivity: a non vanishing {\it vev} of an operator with non trivial
charge is needed, while the density of charge commutes with total charge
operator (which is in fact neutral). We will instead make use of a genuine
disorder parameter that we have constructed and checked on the $U(1)$ gauge
theory\cite{7}.

Another question is if all abelian projections are physically equivalent, a
possibility suggested in\cite{8}. We will study the projection in which $\hat
\Phi$ is the Polyakov line, and the one in which $\hat \Phi$ is any component
of the field strength $G_{\mu\nu}$\cite{8}.

For reasons which will be explained in sect.3, we have technical difficulties
(computing power) to explore the so called maximal abelian gauge\cite{5}.

\section{The disorder parameter: numerical results.}
As for the $U(1)$ case\cite{7} we define the operator which creates a monopole
at the point $\vec z$ and time $z_0$
\begin{equation}
\mu(\vec z,z_0) = {\rm exp}
\left[ \frac{\rm i}{g}\int\,{\rm d}^3y\,f_{0i}(\vec y,z_0) b_i(\vec y-\vec z\,)
\right] \label{eq3.1}\end{equation}
where $f_{0i}$ is the electric field strength  Eq.(\ref{eq2.1}) and 
$b_i/g$ is the vector potential produced by a Dirac monopole, with the Dirac
string subtracted.
$\vec b(\vec r\,)$ is given by
\begin{equation}
\vec b(\vec r) = \frac{\displaystyle \vec r\wedge \vec n}{
\displaystyle r (r - \vec r\cdot\vec n)} \label{eq3.2}\end{equation}
if the gauge is chosen in such a way that the string singularity is in the
direction $\vec n$.
The equal time commutator between the vector potential $a_i$ (Eq.(\ref{eq2.2})
and $f_{0i}$ is
\begin{equation}
\left[a_k(\vec x,x^0),f_{0j}(\vec y,x^0)\right] =
{\rm i} \delta_{kj}\,\delta^3(\vec x - \vec y)
\label{eq3.3}\end{equation}
as in the $U(1)$ gauge theory $f_{0i}$ is the conjugate momentum to $a_i$, and
as a consequence $\mu$ ( Eq.(\ref{eq3.1}) ) is an operator which translates the
field $a_i(x)$ by $b_i(\vec x-\vec z)/g$.

A proper definition of the {\it v.e.v.} of $\mu$ is\cite{7}
\begin{equation}
\langle \mu\rangle =
\frac{\displaystyle
\int{\cal D} A_\mu\,{\rm exp}\left[-\beta S\right]\,\mu(\vec z,z^0)}
{\displaystyle
\int{\cal D} A_\mu\,{\rm exp}\left[-\beta S\right]\,\gamma(z^0)}
\label{eq3.4}\end{equation}
where $\gamma(z^0)$ is a traslation of the field $a_i$ by a time independent
$\vec g(\vec x)$ with $\vec \nabla\wedge\vec g = 0$
\begin{equation}
\gamma(z^0) = {\rm exp}
\left[\frac{\rm i}{g}\int{\rm d}^3x\,f_{0i}(\vec y, z^0) g_i(\vec y)\right]
\label{eq3.5}\end{equation}
subjected to the constraint
\begin{equation}
\int{\rm d}^3x \,{\vec b}^{\,2}(x) = \int{\rm d}^3x \,{\vec g}^{\,2}(x)
\label{eq3.6}\end{equation}
After Wick rotation Eq.(\ref{eq3.4}) can be written
\begin{equation}
\langle \mu\rangle =
\frac{\displaystyle
\int{\cal D} A_\mu\,{\rm exp}\left[-\beta (S+S_b)\right]}
{\displaystyle
\int{\cal D} A_\mu\,{\rm exp}\left[-\beta (S+S_g)\right]}
\label{eq3.7}\end{equation}
with
\begin{mathletters}
\begin{eqnarray}
S_b(\vec x,\vec x^0) &=& \int{\rm d}^3 y b_i(\vec y-\vec x) f_{0i}(\vec y,x^0)\\
S_g(\vec x,\vec x^0) &=& \int{\rm d}^3 y g_i(\vec y) f_{0i}(\vec y,x^0)
\end{eqnarray}
\end{mathletters}
Similarly the correlation function can be defined of any number of monopoles
and antimonopoles: for example for a pair $m$ $\bar m$ at equal time and
distance $d$
\begin{equation}
\langle \mu(\vec d)\mu(0)\rangle =
\frac{\displaystyle
\int{\cal D} A_\mu\,{\rm exp}\left[-\beta (S+S_{b\bar b})\right]}
{\displaystyle
\int{\cal D} A_\mu\,{\rm exp}\left[-\beta (S+S_g)\right]}
\label{eq3.9}\end{equation}
\begin{equation}
S_{b\bar b} = \int{\rm d}^3 y f_{0i}(\vec y,x^0)\left[
b_i(\vec y-\vec d\,) - b_i(\vec y) \right]
\label{eq3.10}\end{equation}
and $g$ is now subjected to the constraint
\begin{equation}
\int{\rm d}^3y \,{\vec g}^{\,2}(y) =
\int{\rm d}^3y \,\left|{\vec b}(\vec y-\vec d\,) - {\vec b}(\vec y)\right|^2
\label{eq3.11}\end{equation}
Instead of $\langle \mu\rangle$ we will measure
\begin{equation}
\rho = \frac{d}{d\beta}\ln\langle\mu\rangle =
\langle S + S_g \rangle_{S+S_g} - \langle S + S_b \rangle_{S+S_b}
\label{eq3.12}\end{equation}
In terms of $\rho$
\begin{equation}
\langle\mu\rangle = {\rm exp}\left[\int_0^\beta\rho(\beta){\rm d}\beta\right]
\label{eq3.13}\end{equation}
For $b\bar b$
\begin{equation}
\rho_{b\bar b} = 
\langle S + S_g \rangle_{S+S_g} - \langle S + S_{b\bar b} 
\rangle_{S+S_{b\bar b}}
\label{eq3.14}\end{equation}
If there is monopole condensation $\langle\mu\rangle\neq0$ or by cluster
property
\begin{equation}
\langle \mu(x)\mu(0)\rangle \mathop\to\limits_{|x|\to \infty}
|\langle\mu\rangle|^2 \neq 0
\label{eq3.17}\end{equation}
In terms of $\rho$ the cluster property Eq.(\ref{eq3.17}) reads
\begin{equation}
\rho_{b\bar b} \mathop\to\limits_{|x|\to \infty} 2\rho_b
\label{eq3.18}\end{equation}

We have measured $\rho$ for a single monopole and for a $m\bar m$ pair at
different distances, across the deconfining phase transition of an $SU(2)$
gauge theory. The abelian projection in the gauge which diagonalizes the
Polyakov line gives a clear signal of condensation: Fig.1 shows $\rho$ for a
$12^3\times 4$ lattice; Fig.2 for a $16^3\times 6$ lattice.A clear
signal is observed of transition from superconductivity to normal
vacuum at $\beta = \beta_c$. The (known) deconfining $\beta_c$ for the
two lattices ($N_T = 4$, $N_T = 6$) is indicated by the vertical lines
in figures 1 and 2.
In Fig.3 $\rho_{b\bar b}$ of a $m\bar m$ pair at  distance $d = 10$
lattice spacing is
compared to $2\cdot\rho_b$, corresponding to a single monopole, checking
successfully Eq.(\ref{eq3.18}).

No signal is observed in the abelian projection which diagonalizes a component
(say $F_{12}$) of the field strength.

A few points about the lattice version of the approach. $f_{0i}$ of 
Eq.(\ref{eq3.1}) is defined by Eq.(\ref{eq2.1}). For the abelian projection in
which $\hat\Phi$ is the direction of log of the Polyakov loop,$L$, the second
term in Eq.(\ref{eq2.1}) is absent when $\mu$ or $\nu$ take the value 0, since
$D_0 L = 0$. Then constructing $f_{0i}$ is simply a projection of $\vec G_{0i}$
on the direction of $L$: of course on the lattice $G_{0i}$ can be taken as the
imaginary part of the plaquette $\Pi_{0i}$.

For the abelian projection in which $\Phi = \ln\Pi_{12}$ the second term of 
 Eq.(\ref{eq2.1}) is not zero, but is computable.

In the case of the so called ``maximal abelian gauge''\cite{5}, in which the
gauge is fixed by maximizing the quantity
\[ M = \sum_{\mu,n}{\rm Tr}\left[ U_\mu(n)\sigma_3
U^\dagger_\mu(n)\sigma_3\right] \]
the effective Higgs $\hat \Phi$ to introduce in Eq.(\ref{eq2.1}) is not known
explicitely, but must be determined by the maximization on each configuration.
This is a serious problem from the numerical point of view, since at each
change of the configuration in the updating procedure to compute $\rho$ by 
Eq.(\ref{eq3.14}) the maximization must be repeated to determine $f_{0i}$ and
$S_b$.
A detailed finite size scaling analysis to extract the thermodinamical limit
from our data is under study.

\section{Conclusions}
We conclude that
\begin{itemize}
\item[1)] gauge theory vacuum is a dual superconductor: the monopoles defined by
the abelian projection diagonalizing the Polyakov line do condense in the
confined phase, and the corresponding dual $U(1)$ symmetry is restored in the
deconfined phase.
\item[2)] Not all abelian projections are physically equivalent: the monopoles
in the gauge in which the field strength is diagonal are irrelevant to
confinement.
\end{itemize}

\vfill\eject
\thispagestyle{empty}
\par\noindent
\epsfxsize0.8\linewidth
{\centerline{\epsfbox{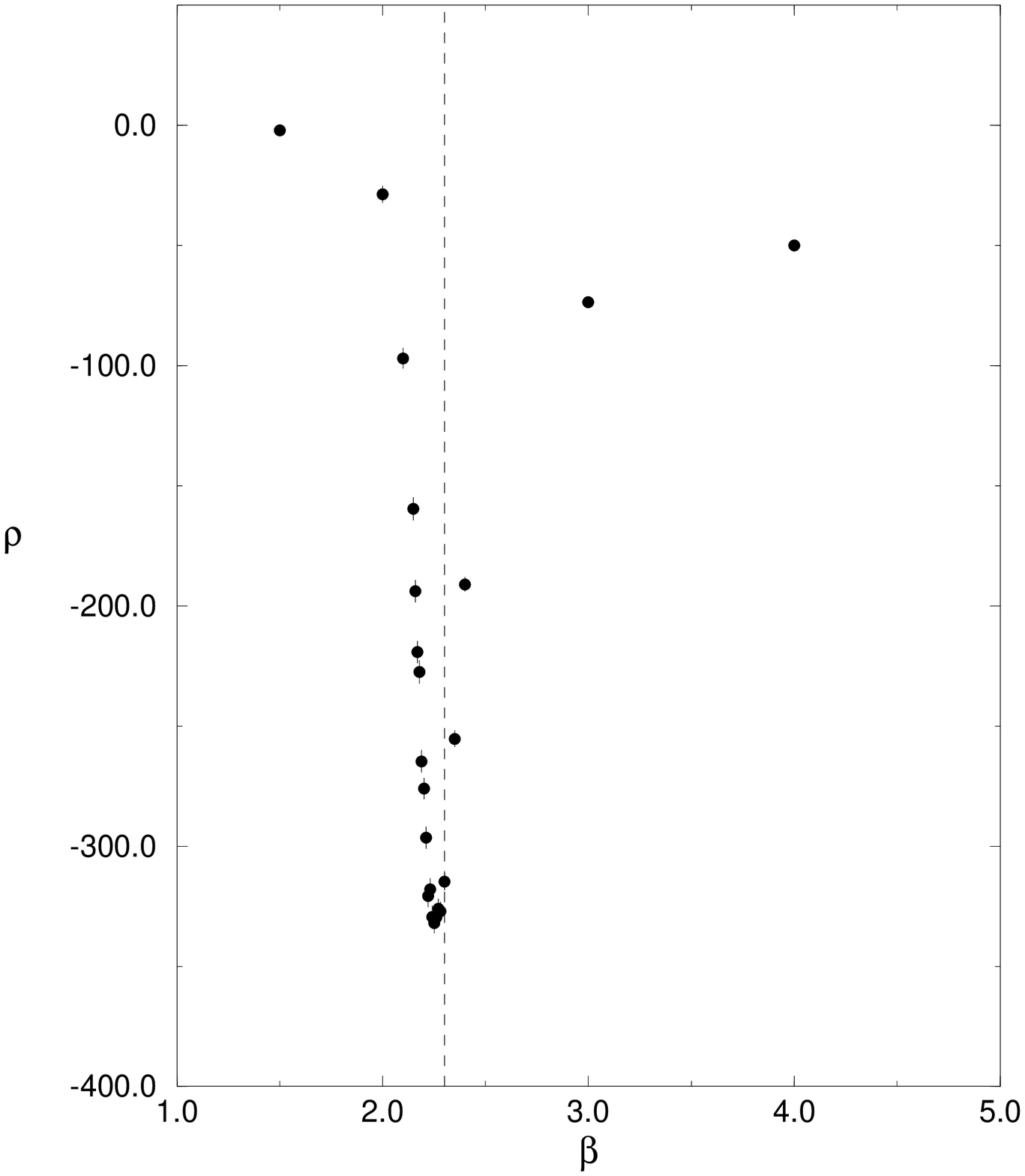}}}
{\centerline{Fig.1}}
\thispagestyle{empty}
\par\noindent
{\centerline{\epsfbox{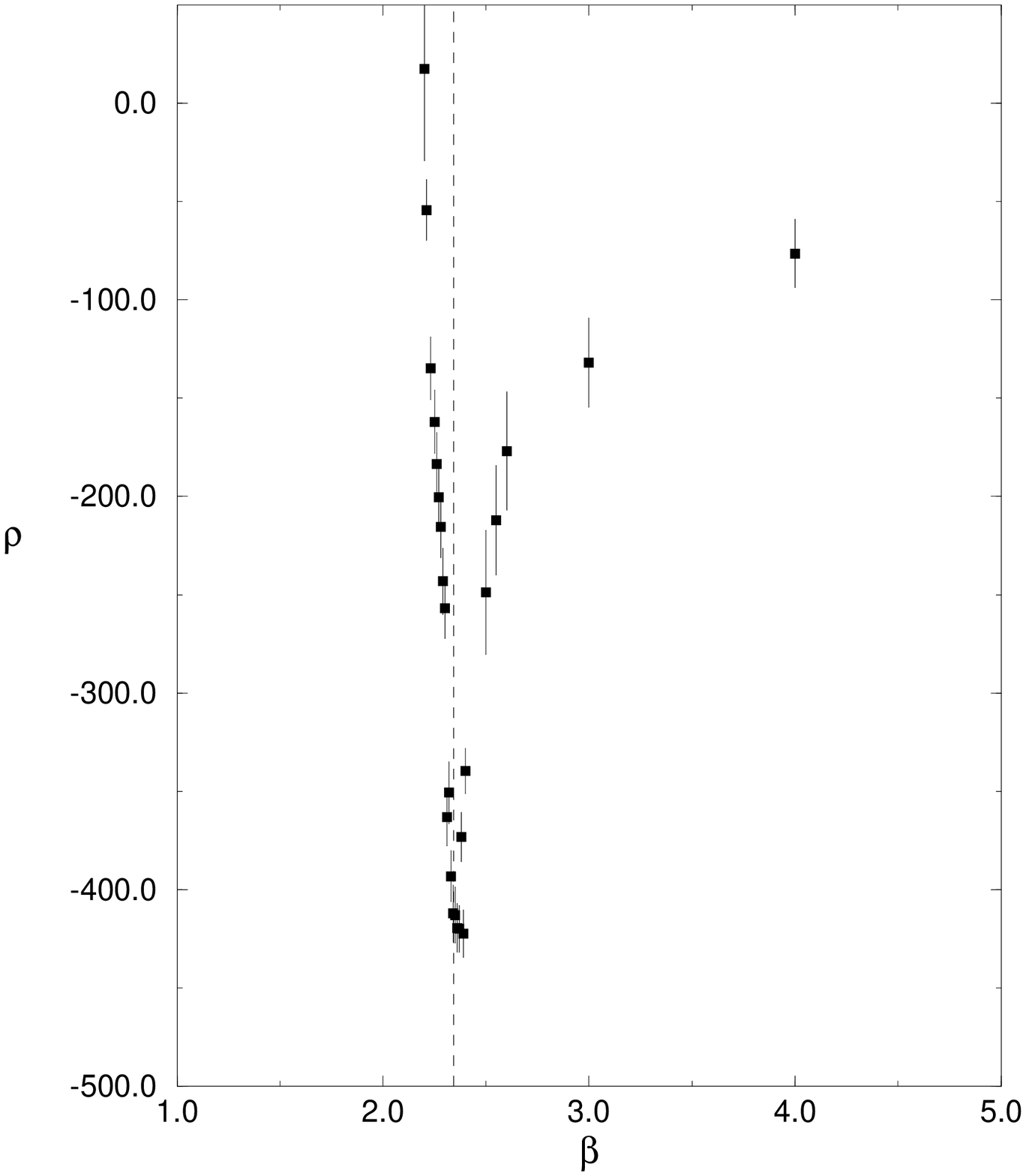}}}
{\centerline{Fig.2}}
\thispagestyle{empty}
\par\noindent
{\centerline{\epsfbox{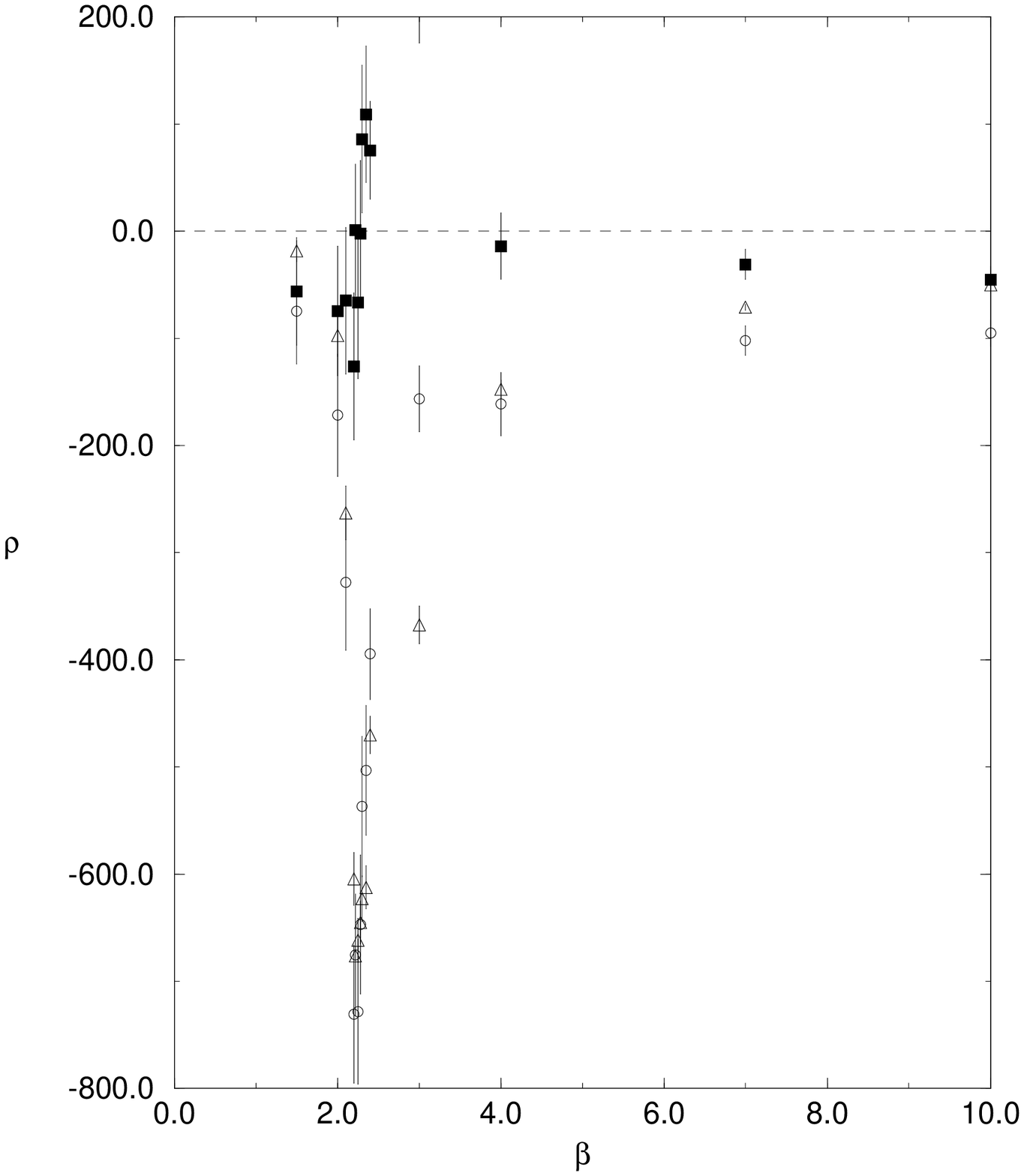}}}
{\centerline{Fig.3}}
\thispagestyle{empty}
\par\noindent
{\centerline{\epsfbox{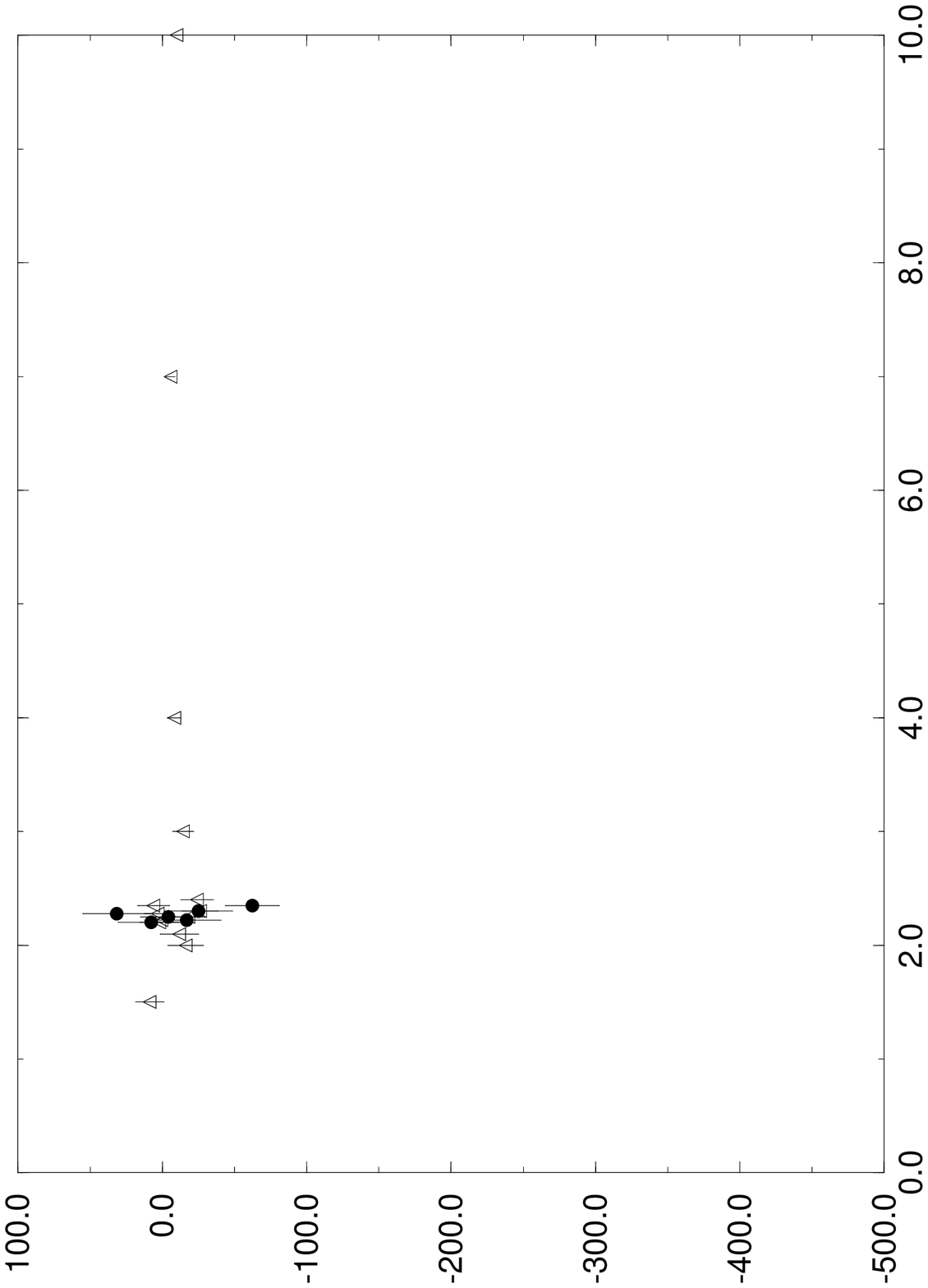}}}
{\centerline{Fig.4}}
\vfill\eject
{\centerline{\Large Figures Captions}}
\vskip0.5truecm
\begin{itemize}
\item[Fig.1] $\rho$ vs $\beta$ on a $12^3\times4$ lattice.
The vertical line denotes $\beta_c$ corresponding to the deconfining transition.
\item[Fig.2] $\rho$ vs $\beta$ on a $16^3\times6$ lattice.
The vertical line denotes $\beta_c$ corresponding to the deconfining
transition.
\item[Fig.3] $\rho_{b\bar b}(d)$ at $d=10$ (circles) compared to
$2\rho$ (triangles) and their difference (squares).
\item[Fig.4] $\rho$ for the abelian projection diagonalising $F_{12}$
on $8^3\times4$ (dots) and $12^3\times4$ (triangles) lattices.
\end{itemize}
\end{document}